# Translating Chirality into Multidirectional Motion through Broadband Chiroptical MXenes

*Wookjin Jung[1], Dongkyu Lee[1], Yonghee Lee[2], Ki Hyun Park[1], and Jihyeon Yeom[1,3,4,5*].*

[1] Department of Materials Science and Engineering, Korea Advanced Institute of Science and Technology (KAIST), 291 Daehak-ro, Yuseong-gu, Daejeon 34141, Republic of Korea

[2] Department of Nano & Advanced Materials Science and Engineering, Kyungpook National University (KNU), 2559 Gyeongsang-daero, Sangju-si, Gyeongsangbuk-do 37224, Republic of Korea

[3] Department of Biological Sciences, Korea Advanced Institute of Science and Technology (KAIST), 291 Daehak-ro, Yuseong-gu, Daejeon 34141, Republic of Korea

[4] Institute for Health Science and Technology, Korea Advanced Institute of Science and Technology (KAIST), 291 Daehak-ro, Yuseong-gu, Daejeon 34141, Republic of Korea

[5] Institute for the NanoCentury, Korea Advanced Institute of Science and Technology (KAIST), 291 Daehak-ro, Yuseong-gu, Daejeon 34141, Republic of Korea

[*] Corresponding author: jhyeom@kaist.ac.kr

**ABSTRACT:**

The integration of chirality into functional materials enables control of light–matter interactions beyond binary illumination (on/off). Conventional photoactuators rely on binary modulation, limiting them to unidirectional motion. In contrast, we introduce a ternary optical logic paradigm where actuation direction is encoded by the handedness of circularly polarized light (CPL). Here, we establish a chiral $Ti_3C_2T_x$ MXene platform bridging molecular chirality and mechanical actuation. Phenylalanine enantiomers are covalently anchored onto MXene nanoflakes via chiral nanopainting. The 2D confinement forces ligands into vertically aligned supramolecular networks. Interlayer-spacing analysis and simulations corroborate that such supramolecular networks unlock exceptionally broadband circular dichroism from the ultraviolet to the near-infrared. This supramolecular chirality synergizes with MXene's plasmonic properties to drive handedness-dependent photothermal conversion, with a 30% differential temperature rise between matched and mismatched CPL. Embedding this chiral MXene into thermoresponsive hydrogels realizes, to the best of our knowledge, the first CPL-driven soft actuator that implements LCP/RCP/off as a ternary input to program multidirectional deformation based on a photothermal mechanism. This molecular-to-macroscopic translation demonstrates a new paradigm for chirality-encoded soft robotics and adaptive photonics.

Chirality—the property that distinguishes an object from its mirror image—underpins asymmetric interactions across chemistry, biology, and condensed matter physics. At the molecular level, chirality dictates enantioselective recognition and reaction kinetics; at the mesoscale, it determines supramolecular organization; and at the macroscopic level, it can encode handedness into bulk optical and mechanical behavior. Beyond its well-known biochemical role, recent advances have demonstrated that chiral organization can modulate magnetic, electronic, and photonic phenomena, establishing chirality as an active design variable for functional materials rather than a mere structural descriptor.[1-9]

Among the manifestations of chirality, circularly polarized light (CPL) represents the optical counterpart of molecular handedness. The helical rotation of its electric and magnetic field vectors introduces an intrinsic sense of directionality that can be selectively coupled to chiral matter. Conventional light–matter interactions with unpolarized light are binary—governed only by illumination on/off—whereas CPL introduces handedness as a third logical state. This additional degree of freedom provides the conceptual foundation for ternary light-matter systems, in which left-handed, right-handed, and light-off inputs elicit three distinct material responses.[10-13] Such CPL-encoded logic could revolutionize photonic actuation, adaptive optics, and bio-robotic control if translated into macroscopic, reversible deformation.

To realize this vision, a material must meet two stringent criteria:

(i) it must exhibit optical activity, responding differentially to left- and right-handed CPL, and

(ii) it must be photoresponsive, transducing absorbed optical energy into thermal, electrical, or mechanical outputs.[14-17]

Two-dimensional (2D) materials provide an ideal foundation for such coupling owing to their ultrathin geometry, anisotropic electronic structure, and rich surface chemistry. Their large surface-to-volume ratios amplify the effects of surface functionalization, allowing molecular chirality to be efficiently translated into collective optical anisotropy.[18-20]

However, current chiral 2D systems remain far from fulfilling the requirements for CPL-driven actuation. Graphene-based hybrids possess high conductivity but have chemically inert basal planes, yielding only weak optical activity limited to the ultraviolet region.[21-24] Chiral MXenes, although structurally versatile and hydrophilic, have so far exhibited narrowband circular dichroism with only marginal CPL selectivity beyond ~400 nm.[20, 25-28] Attempts to extend their response using helicene derivatives have improved the spectral range but introduced issues of poor solubility and weak chiroptical coupling. Consequently, the creation of broad-spectrum, solution-processable, and CPL-responsive chiral MXene materials remains a critical unmet goal.

Equally important is the functional translation of CPL selectivity into controllable macroscopic motion. Existing light-driven soft actuators—based on photothermal, photochemical, or photoisomerization mechanisms—typically rely on intensity modulation or wavelength switching as control parameters.[29-34] These systems are inherently unidirectional: they bend or contract along a single predefined axis under illumination and relax upon cooling. Introducing chirality into light-responsive materials opens a qualitatively new control paradigm: the ability to dictate actuation direction by the handedness of light, rather than its power or wavelength. Such multidirectional actuation is a crucial step toward chirality-encoded soft robotics and adaptive architectures that can autonomously reconfigure their shape in response to optical stimuli. Yet, despite theoretical predictions, the experimental realization of reversible, CPL-handedness-dependent deformation with programmable geometry has remained elusive.

Here, we report a broadband chiroptical MXene platform that bridges this gap. Using a chiral nanopainting strategy, we covalently anchored phenylalanine enantiomers onto $Ti_3C_2T_x$ MXene nanoflakes, forming supramolecularly ordered aromatic networks that impart strong circular dichroism from the ultraviolet through the near-infrared region. Unlike zero-dimensional nanoparticles where surface curvature limits ligand ordering, the atomically flat topology of 2D MXene imposes a distinct steric constraint that induces long-range vertical alignment of chiral ligands, thereby maximizing chiroptical coupling efficiency. This supramolecular chirality couples synergistically with the intrinsic plasmonic and photothermal properties of MXene, enabling handedness-dependent photothermal conversion—up to 30% higher under matched CPL. When embedded into a thermoresponsive hydrogel, the chiral MXene transduces molecular asymmetry into macroscopic motion, yielding a CPL-responsive soft actuator capable of reversible upward, downward, and neutral (flattened) states corresponding to LCP, RCP, and CPL-off inputs.

Extending this principle, we designed a CPL-programmable hydrogel architecture that spatially encodes chiral MXenes of opposite handedness, producing complex, multidirectional deformations that can be reconfigured by localized CPL irradiation. Unlike traditional soft actuators that prioritize speed or force, our work presents a CPL-driven soft-matter platform that realizes ternary optomechanical logic based solely on light polarization. This validates a new control dimension for soft robotics, where instructional complexity is embedded in the light's handedness. This multiscale coupling—from molecular chirality to supramolecular ordering, optoelectronic asymmetry, and mechanical actuation—establishes a general strategy for chirality-encoded soft robotics, photonic logic, and bioresponsive devices based on chiral 2D materials.

**Results and Discussion**

We synthesized chiral MXene nanoflakes by adapting the previously reported chiral nanopainting strategy to achiral $Ti_3C_2T_x$ (Scheme 1).[5] Briefly, achiral $Ti_3C_2T_x$ nanoflakes were obtained via intercalation of bulk $Ti_3AlC_2$. The resulting $Ti_3C_2T_x$ MXene nanoflakes were surface-treated with (3-aminopropyl)triethoxysilane (APTES) to anchor chiral amino acids on their surface and to protect them from undesirable chemical reactions such as oxidation.[35-37] Here, we rationally selected phenylalanine as the chiral surface ligand to promote supramolecular alignment through aromatic π–π interactions facilitated by the atomically flat 2D basal planes of MXene. Subsequently, *L*- or *D*-phenylalanine was covalently conjugated to the APTES-treated MXene nanoflakes to impart chirality to the surface layer.

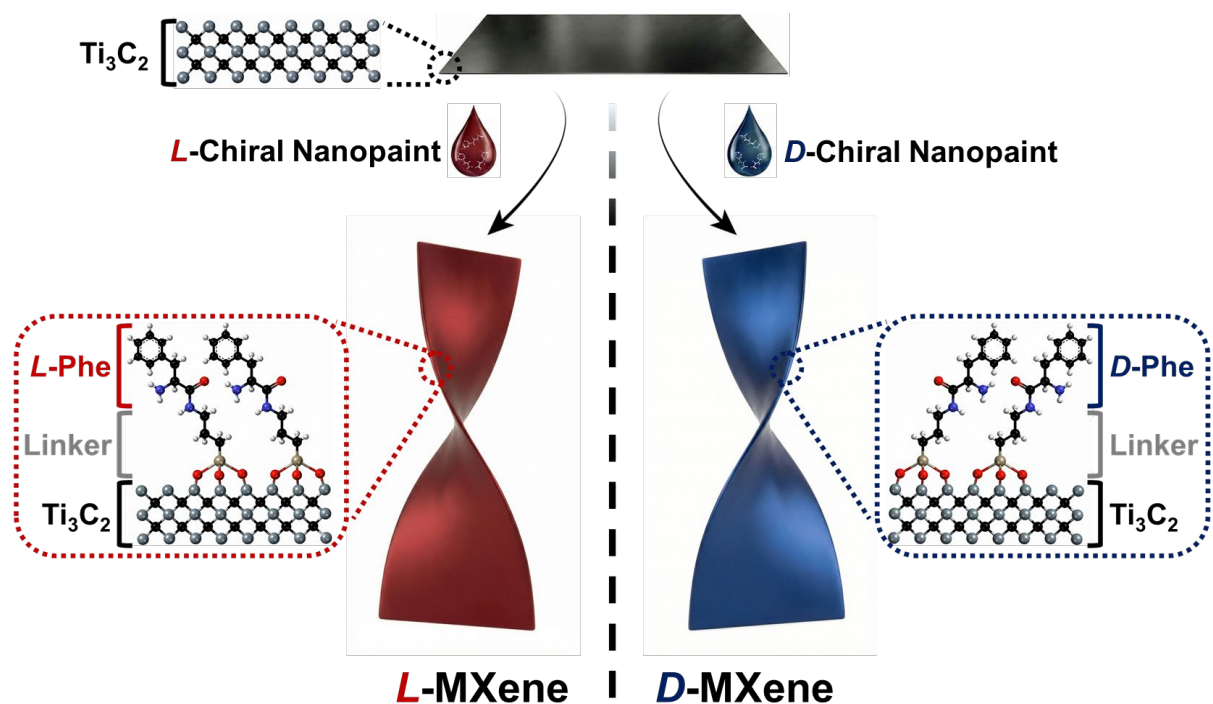

**Scheme 1.** Synthesis and structure of chiral MXene nanoflakes using a chiral nanopainting method.

To evaluate whether the synthesized MXene nanoflakes treated with the chiral nanopaint were optically active, circular dichroism (CD) spectra of the aqueous dispersion of chiral MXene were obtained **(Figure 1A, B, S1A)**. While the MXene nanoflakes with chiral nanopaint (*L*- and *D*-MXene) showed mirror-imaged CD signals, MXene nanoflakes with racemic nanopaint (*DL*-MXene), pristine Ti$_3$C$_2$T$_x$ MXene and MXene nanoflakes treated only with achiral APTES showed no significant optical activity **(Figure S1B-D)**. Interestingly, the synthesized chiral MXene nanoflakes exhibited optical activity across a broad range spanning the UV, VIS, and NIR regions.

Exhibiting broad-range optical activity, the chiral MXene nanoflakes retained their characteristic 2D morphology, as confirmed by transmission electron microscopy (TEM) and scanning electron microscopy (SEM) **(Figure 1C, D)**. We further examined the vertical cross-section of the chiral MXene using TEM imaging **(Figure S2A, B)**. As expected, the characteristic layered structure typical of 2D nanomaterials including MXene was clearly observed. Notably, analysis of the TEM images revealed that the individual interlayer distance of the chiral MXene nanoflakes was 1.810 nm **(Figure S2C)**. Considering that the typical interlayer distance of Ti$_3$C$_2$T$_x$ MXene is around 1.0 nm, this increase in interlayer distance arose from the chiral nanopaint coating introduced during the synthesis of the chiral MXene.[38, 39]

The presence of the chiral nanopaint layer was further supported by energy-dispersive X-ray spectroscopy (EDS) analysis of the chiral MXene **(Figure 1E)**. In addition to the Ti and F elements typically observed in Ti$_3$C$_2$T$_x$ MXene, the EDS spectra revealed the presence of Si and N elements, which were not intrinsic to the MXene itself but characteristic of the chiral nanopaint, thereby confirming the successful surface functionalization.

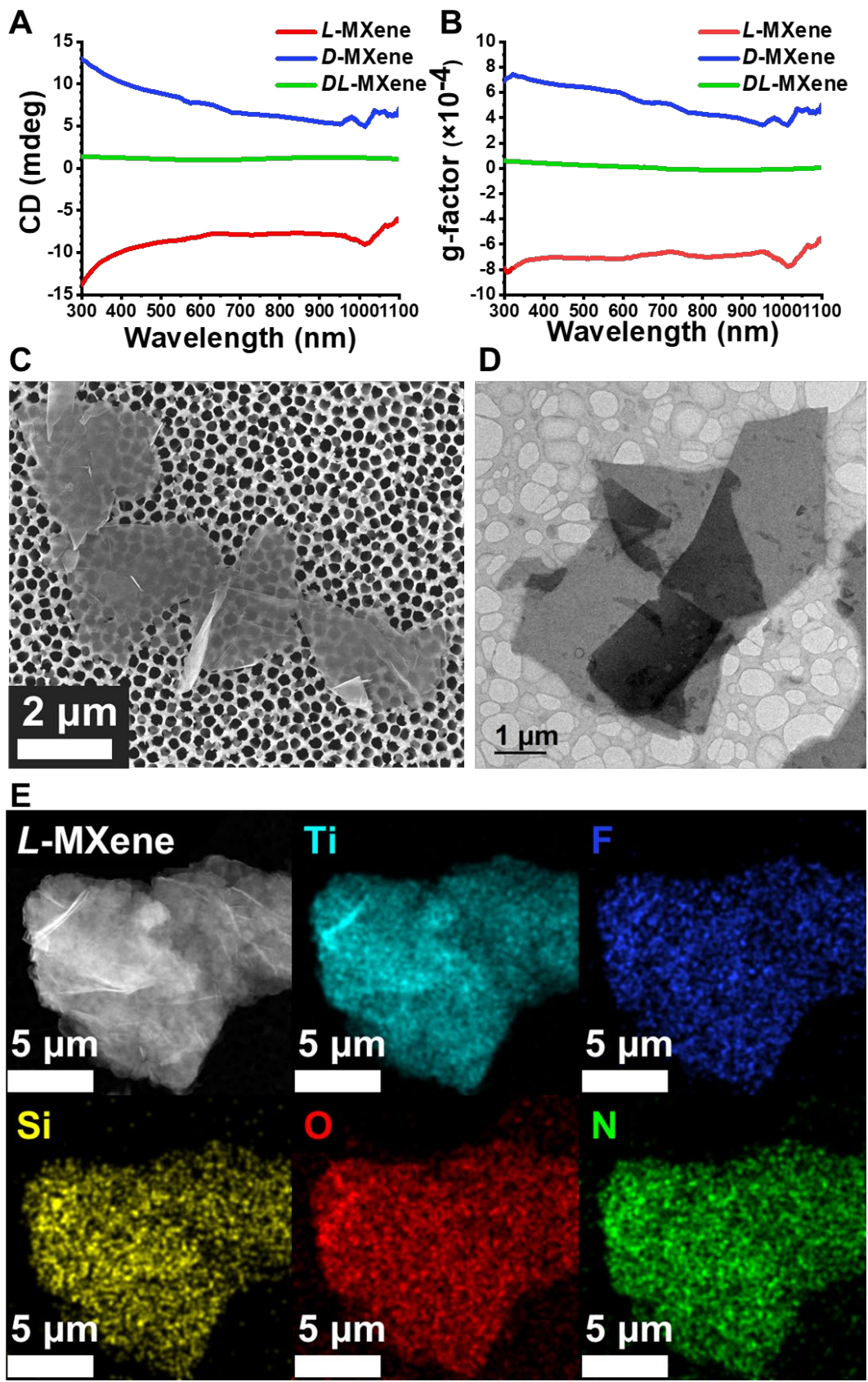

**Figure 1.** Synthesized chiral MXene nanoflakes. (A) CD and (B) g-factor spectra of chiral MXene in aqueous dispersions. (C) SEM and (D) TEM images of *L*-MXene nanoflakes. (E) A scanning transmission electron microscopy image of the *L*-MXene nanoflake and its EDS analysis results (Ti for light-blue, F for navy, Si for yellow, O for red, and N for green).

X-ray photoelectron spectroscopy (XPS) analysis of pristine MXene, APTES-MXene, and chiral MXene (*L*-MXene) was performed to study the chemical compositions and oxidation states **(Figure S3)**. The pristine MXene showed the presence of TiC, Ti-X, $Ti_xO_y$ and $TiO_2$ peaks. The C 1s spectra were fitted with C-Ti, C-Ti-O, C-C, C-O, and C-F peaks, and the O 1s peaks, composed of C-Ti-O, Ti-O-Ti, and C-O, were well matched with those of $Ti_3C_2T_x$ MXene.[40] In N 1s spectra, no visible peaks were observed. After surface modification using APTES, similar Ti 2p spectra and associated components were observed. However, C-N, C-Si, and Si-O peaks emerged at 286.2 eV, 283.8 eV, and 532.4 eV in C 1s and O 1s spectra, respectively, which suggested the successful covalent surface modification using APTES.[41] The existence of $NH_2$ and $NH_3$ peaks in N 1s spectra corroborated the results.[42] When phenylalanine molecules were conjugated with APTES-MXene, amide bonds (HNC=O) were formed. Here, the HNC=O peak emerged at 288.0 eV.[43] In O 1s and N 1s spectra, corresponding peaks of HNC=O bonds were also detected. These results collectively indicated that the chiral nanopaint was firmly anchored onto the surface of the $Ti_3C_2T_x$ MXene through covalent bonding.

To investigate the origin of the broad-range optical activity observed in the chiral MXene, we first compared the CD spectra of the aqueous dispersions of chiral MXene with those of aqueous solutions of enantiomeric phenylalanine **(Figure S4)**. Considering that the aqueous solutions of phenylalanine exhibited optical activity only in the narrow UV range, it was confirmed that the broad optical activity of chiral MXene cannot be attributed solely to the individual phenylalanine molecules anchored on its surface.

When molecules bearing hydrophobic aromatic moieties, such as phenylalanine, are anchored on hydrophilic nanostructured surfaces in aqueous media, they tend to adopt an upright orientation with respect to the substrate.[44, 45] This preferential orientation is driven by supramolecular interactions among the aromatic groups.[46-48] Therefore, we hypothesized that the unsubstituted benzene rings drive the anchored phenylalanine molecules to adopt an upright orientation relative to the MXene surface, stabilized by intermolecular aromatic interactions. We further reasoned that this chiral supramolecular organization would make a substantial contribution to the observed optical activity. To test this hypothesis, we focused on the interlayer distance of the chiral MXene. The interlayer distance measured for our chiral MXene was ~1.81 nm, which corresponded to an increase of 8.1 Å compared to that of untreated $Ti_3C_2T_x$ MXene (~1.0 nm). When $Ti_3C_2T_x$ MXene was modified solely with an aminosilane coupling agent such as APTES, the interlayer spacing increased by only ~1.5 Å.[49, 50] Therefore, the aromatic moieties at the terminal ends of the chiral MXene accounted for an additional ~6.6 Å increase in the interlayer spacing. Considering the geometric dimensions of these aromatic moieties, an upright orientation with respect to the MXene nanoflake surface was expected to produce an increase of ~6.5 Å in the interlayer spacing, whereas a flat-lying orientation would result in an increase of only ~3.5 Å.[51-55] These geometric analyses collectively suggest that the aromatic rings at the chiral MXene surface adopted an upright-oriented geometry stabilized by supramolecular arrangements among neighboring aromatic rings within the interlayer gallery. These supramolecular arrangements induced collective chiral organization at the MXene surface, enabling the emergence of the pronounced optical activity observed in the chiral MXene.

When the hydrogen atoms on the benzene ring of phenylalanine are substituted with hydroxyl groups, the resulting hydroxylated amino acids adopt less ordered arrangements in aqueous solution than phenylalanine, owing to enhanced hydrogen bonding interactions.[56, 57] Consequently, we anticipated that MXene nanoflakes decorated with such hydroxylated amino acids would exhibit a reduced degree of chirality compared with those functionalized with phenylalanine. To test this hypothesis, we synthesized and compared MXene nanoflakes using different chiral amino acids, tyrosine (Tyr) and 3,4-dihydroxyphenylalanine (DOPA), instead of phenylalanine (Phe) for surface modification **(Figure S5)**. While *L*-Tyr and *L*-DOPA-modified MXene showed negligible CD and g-factor values, *L*-MXene exhibited considerable and consistent g-factor values across the UV-NIR range, despite the structural similarities among Phe, Tyr, and DOPA molecules. These findings suggested that the optical activity of the chiral MXene arose from the unsubstituted benzene rings of the phenylalanine molecules located on their surface.

To corroborate the pivotal role of unsubstituted benzene rings in the supramolecular ordering of surface-anchored phenylalanine, we conducted automated docking simulations using AutoDock Vina with a gradient-based genetic algorithm.[58, 59] Specifically, the center of the benzene ring of each *L*-enantiomeric amino acid was positioned at the origin of the coordinate space, and an identical amino acid molecule was allowed to approach the reference molecule to simulate intermolecular interactions. For each amino acid type, the top 50 intermolecular configurations, ranked by binding affinity, were extracted. From these docking results, we calculated the corresponding intermolecular affinity values, the relative positions of the benzene ring centers, and the normal vectors of the aromatic rings. As a result, the median intermolecular affinity between two *L*-Phe molecules was calculated to be −5.44 kcal mol$^{−1}$, which was 3.69-fold stronger than that between two *L*-Tyr molecules and 2.85-fold stronger than that between two *L*-DOPA molecules **(Figure 2A)**. In addition, the median distance between the benzene rings of two *L*-Phe molecules was 3.95 Å, corresponding to only 82.5% and 80.9% of the respective median distances observed for *L*-Tyr and *L*-DOPA **(Figure 2B)**. These results indicated that *L*-Phe molecules exhibited stronger intermolecular affinity and adopted a more densely packed spatial arrangement compared to *L*-Tyr or *L*-DOPA molecules.

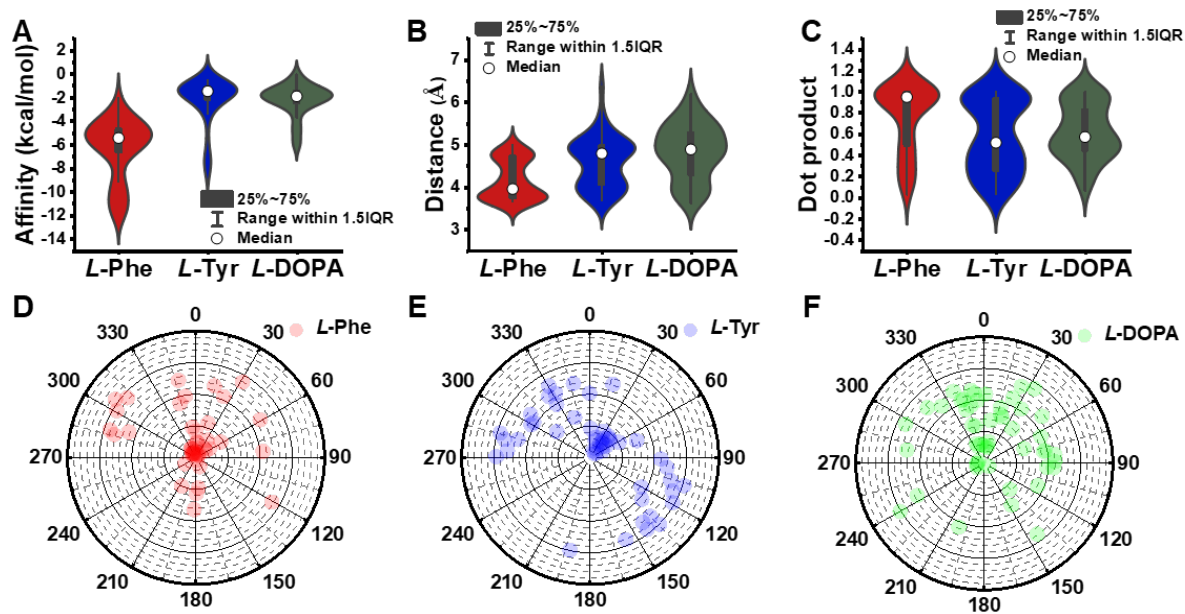

**Figure 2.** Molecular docking simulation of pairs of *L*-Phe, *L*-Tyr and *L*-DOPA molecules. (A) Intermolecular affinity of each pair of amino acid molecules. (B) Distance between centers of two benzene rings of each pair of amino acid molecules. (C) Dot product between orientation vectors of two benzene rings of each pair of amino acid molecules. Distribution of orientation vectors of the benzene rings within single (D) *L*-Phe, (E) *L*-Tyr, and (F) *L*-DOPA molecules when an additional identical molecule was docked to the reference molecule located at the origin of the coordinate space.

To quantify the spatial orientation of the approaching amino acid molecules, we calculated the dot product of the normal vectors of the benzene rings from the reference and approaching amino acids **(Figure 2C)**. A value closer to 1 indicated that the spatial orientation of the benzene rings in the two amino acid molecules approached a highly ordered, parallel alignment. For *L*-Phe, this value reached 0.947, remarkably close to the maximum of 1, whereas the corresponding values for *L*-Tyr and *L*-DOPA were only 0.517 and 0.571, respectively.

By analyzing the spatial distribution of the normal vectors of the benzene rings in amino acid molecules approaching the surface, we further confirmed that the unsubstituted benzene rings in phenylalanine played a key role in establishing the highly ordered supramolecular arrangement among phenylalanine molecules **(Figure 2D-F)**. The results revealed a progressively stronger clustering tendency around the coordinate center in the order of *L*-DOPA, *L*-Tyr, and *L*-Phe. The orientations of the normal vectors indicated that multiple identical amino acid molecules adopted increasingly ordered arrangements in the same order.

Taken together, these results indicated that the unsubstituted benzene rings played a pivotal role in enabling the anchored phenylalanine molecules to form a highly ordered supramolecular arrangement. This ordered supramolecular organization provided the structural basis for the broad chiroptical activity observed in the chiral MXene.

We next investigated how disrupting the intermolecular interactions among these unsubstituted benzene rings would influence the chiroptical properties of chiral MXene. Specifically, *L*- and *D*-MXene were dispersed at equal concentrations in either deionized water or dimethyl sulfoxide (DMSO), and their CD spectra were measured **(Figure S6)**. The chiral MXene dispersions in water exhibited pronounced optical activity extending from the UV region to the NIR region. In sharp contrast, when dispersed in DMSO at the same concentration, the chiral MXene displayed markedly reduced CD intensities. Notably, for MXene dispersions in DMSO, the CD signals became negligible above 550 nm. Since DMSO is aprotic and has lower polarity compared to water, it interferes with supramolecular interactions involving aromatic rings.[60-63] Therefore, the markedly reduced optical activity strongly suggested that the broad optical activity of chiral MXene in aqueous dispersion originated from supramolecular interactions mediated by the unsubstituted benzene rings of Phe molecules.

We also prepared the chiral MXene in the form of dried films and measured the CD spectra of these films **(Figure S7)**. As a result, they lost their optical activity in the solid film state. Upon drying, the CD signal of the chiral MXene diminished to baseline across the entire spectral range, indicating that supramolecular chirality was maintained only in the hydrated, dispersed state. Taken together, these results indicate that the loss of optical activity upon drying arose from the collapse of supramolecular chirality, caused by disruption of the spatial orientation of Phe molecules. In aqueous dispersion, the Phe molecules anchored on the chiral

MXene surface adopted a well-defined spatial orientation. This spatial orientation of the molecules dictated the chiral morphology of the MXene nanoflakes in the aqueous state. Such structural chirality further imparted spatial anisotropy to the surface plasmonic resonance and free carrier oscillations within the nanoflakes.[64-67] As a result, the chiral MXene in the aqueous phase exhibited pronounced optical activity by preferentially absorbing CPL of a specific handedness.

The intermolecular interactions among Phe molecules anchored to the MXene nanoflake surface endowed the chiral MXene with broad optical activity in aqueous dispersion. It should be noted that the synthesized materials retain the fundamental optoelectronic merits of MXenes. MXene inherently exhibits localized surface plasmon resonance (LSPR) upon light irradiation due to its high carrier density and excellent electrical conductivity.[68, 69] When incident photons couple with surface plasmons in MXene, part of the photon energy undergoes non-radiative relaxation via electron transitions, which subsequently induces lattice vibrations and enables highly efficient photothermal conversion.[70] This implies that, upon irradiation of chiral MXene with CPL, the handedness of the incident CPL dictates differences in plasmon-driven optoelectronic interactions, even when the chiral MXene shares the same handedness. Based on this mechanism, we hypothesized that the optical activity of the chiral MXene could be exploited so that chiral MXene with a specific handedness would exhibit higher photothermal conversion efficiency under CPL of the matching handedness compared to CPL of the opposite handedness.

To test this hypothesis, we irradiated identical aqueous chiral MXene dispersions with either left-handed (LCP) or right-handed (RCP) circularly polarized light at 780 nm in the NIR region and monitored the resulting temperature changes. The wavelength of 780 nm was selected to validate the CPL responsiveness of the photothermal conversion system in the NIR region. This spectral window is known for its high penetration through biological tissues, thereby rendering the chiral MXene particularly suitable for further applications.[71, 72] After 15 min of irradiation, the *L*-MXene dispersion exposed to RCP exhibited a temperature rise of 40.5 °C, whereas that exposed to LCP showed a rise of 31.3 °C **(Figure S8A)**. When *D*-MXene dispersions were irradiated with LCP or RCP, the dispersion exposed to RCP exhibited a temperature rise of 29.2 °C, whereas that exposed to LCP showed a rise of 38.2 °C **(Figure S8B)**. These results demonstrated that the temperature increase differed by approximately 30% depending on the handedness of the incident CPL. This phenomenon was due to the optical activity of chiral MXene. For instance, in our system, the CD signal of *L*-MXene was consistently negative across the 300-1100 nm range, as shown in Figure 1A. This negative value indicated that the dispersion of *L*-MXene absorbed RCP more strongly than LCP throughout this spectral region.[73] This confirmed that the observed photothermal differences were directly linked to the handedness-selective absorption shown in the CD spectra.

Given the combination of its pronounced optical activity and the intrinsic optoelectronic properties of MXene, our chiral MXene displayed substantial differences in photothermal conversion efficiency depending on the handedness of the incident CPL. However, while these results demonstrated the CPL-responsiveness of the chiral MXene, they were insufficient to establish a ternary light-matter interaction system based on LCP, RCP, and the CPL-off state. To address this limitation, we designed a CPL-responsive actuator whose direction of actuation was controlled by the handedness of incident CPL. Specifically, we envisioned that dispersing the chiral MXene within a thermoresponsive poly(*N*-isopropylacrylamide) (PNIPAM)-based hydrogel would enable CPL-triggered deformation through photothermal heating.[74] Upon CPL irradiation, the embedded chiral MXene would generate heat via photothermal conversion, inducing contraction of the hydrogel based on the thermoresponsiveness of the PNIPAM hydrogel. Importantly, this this differential temperature response provided the basis for chirality-encoded discrimination, which is a prerequisite for ternary logic systems in which three distinct material states must be reliably and reproducibly accessed using optical input alone. In particular, the observed ~30% difference provided sufficient thermal contrast to drive distinct conformational changes in thermoresponsive polymers, thereby enabling CPL handedness to serve as a genuine ternary logical input rather than a simple efficiency modifier.

To validate this concept, we fabricated PNIPAM-based hydrogels containing chiral MXene. The resulting constructs were confirmed to be hydrogels, as they retained their structure without flowing when they were inverted **(Figure 3A)**. We then examined whether the chiroptical properties of chiral MXene were preserved after incorporation into the hydrogel network by measuring CD spectra **(Figure 3B, C)**. As observed for aqueous dispersions, *L*-MXene hydrogels exhibited negative CD signals across a broad spectral range, whereas *D*-MXene hydrogels displayed positive CD signals, confirming the retention of chiroptical activity within the hydrogel network.

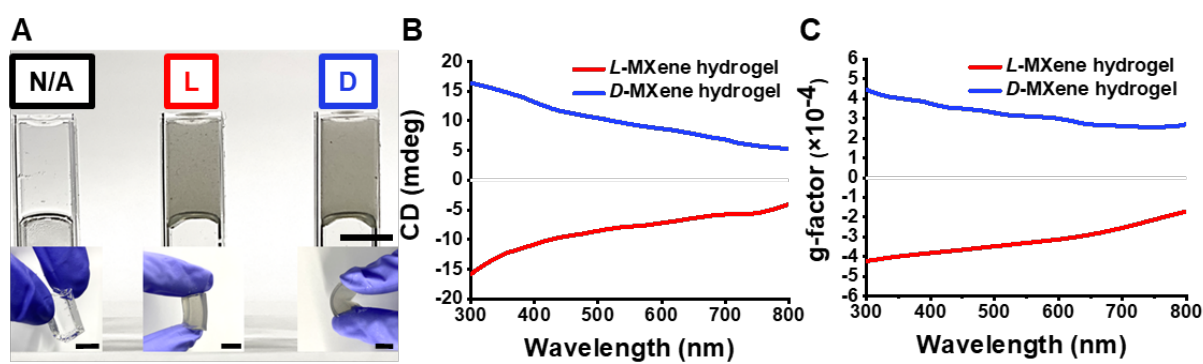

**Figure 3.** PNIPAM-based chiral hydrogel with chiral MXene. (A) Photograph of pure PNIPAM hydrogel and the chiral hydrogels with different handedness. The scale bars are 1 cm. (B) CD spectra and (C) g-factor spectra of the chiral hydrogels, showing their optical activity depending on the handedness of the embedded chiral MXene.

As the optical activity of chiral MXene was maintained in the hydrogel network, we next designed a CPL-responsive actuator based on asymmetric placement of *L*- and *D*-MXene. To achieve multidirectional actuation, the hydrogel matrix was designed with a multilayered structure. The top layer contained *D*-MXene, exhibiting stronger photothermal conversion under LCP irradiation, and the bottom layer contained *L*-MXene, exhibiting stronger photothermal conversion under RCP irradiation **(Figure 4A)**. Under RCP irradiation at the center of the actuator, the *L*-MXene-containing bottom layer would generate greater heat and undergo stronger contraction relative to the *D*-MXene-containing top layer, resulting in a net downward bending motion. Conversely, under LCP irradiation, the top *D*-MXene layer would contract more strongly than the bottom *L*-MXene layer, leading to upward bending of the actuator.

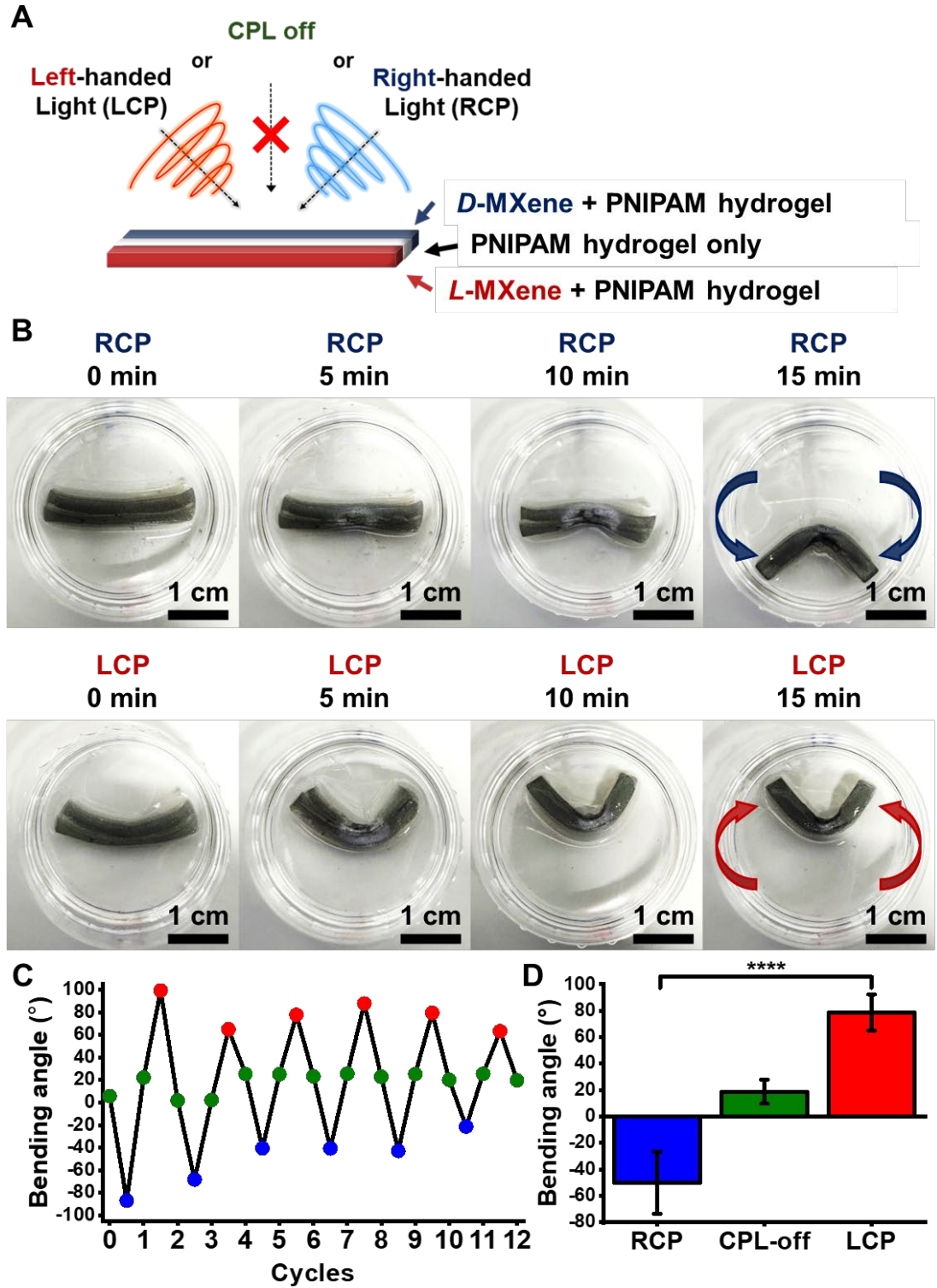

**Figure 4.** CPL-responsive soft actuator using PNIPAM-based hydrogel with chiral MXene. (A) Schematic illustration of the CPL-responsive soft actuator. (B) Photographs showing reversible bending of the actuator under alternating LCP and RCP irradiation. Arrows indicate the direction of bending under corresponding irradiation of CPL. (C) Repeated actuation cycles of the CPL-responsive soft actuator under alternating RCP and LCP irradiation with cooling steps. The plotted bending angle corresponds to the central angle of the actuator. Red dots indicate the upward bending through LCP irradiation, blue dots indicate the downward bending through RCP irradiation, and green dots indicate the flattening of the actuator through CPL-off. (D) Bending angle of the actuator under different CPL inputs. Error bars indicate the standard deviation. ****$p < 0.0001$, calculated by one-way analysis of variance (ANOVA).

To experimentally validate this hypothesis, we fabricated a multilayer actuator by positioning a PNIPAM hydrogel layer containing *L*-MXene at the bottom, a *D*-MXene-containing layer at the top, and a thin achiral PNIPAM spacer in between. The top and bottom layers were prepared with equal concentrations of chiral MXene, distinguished solely by their handedness, as validated by optical density measurements. Then, the actuator was irradiated with RCP and LCP light. Upon 15 min of RCP irradiation, the actuator bent downward as hypothesized. When the light was turned off and the actuator was exposed to ambient temperature, it flattened toward its original state. Subsequent irradiation with LCP for 15 min induced upward bending, confirming the enantioselective actuation behavior **(Figure 4B)**. By contrast, when both the top and bottom layers contained racemic *DL*-MXene, 15 min of CPL irradiation led only to symmetric contraction arising from a symmetric temperature increase at the actuator center, without any detectable bending toward a particular direction **(Figure S9A)**. Likewise, even for hydrogel actuators incorporating the *L*-MXene and the *D*-MXene, illumination with linearly polarized light instead of CPL produced only symmetric, thermally induced contraction at the center, with no preferential bending direction **(Figure S9B)**. These control experiments demonstrated that directional bending in our CPL-responsive actuators arose solely from the combination of the actuator's built-in chiral asymmetry and the handedness of the incident CPL.

An actuation cycle consisting of 15 min CPL irradiation and cooling was repeated ten additional times, and in each case the bending direction was consistently governed by the handedness of the incident CPL **(Figure 4C, S10)**. These results clearly demonstrated the reversible and multidirectional movement of the actuator. During repeated actuation cycles, irradiation with RCP yielded an average bending angle of −50.1°, corresponding to downward bending, whereas in the CPL-off state the actuator relaxed to a relatively flattened configuration with an average bending angle of 18.8°. Subsequent LCP irradiation produced an average bending angle of 78.7°, corresponding to upward bending **(Figure 4D)**. This baseline offset in the CPL-off state was attributed to intrinsic structural asymmetry, such as meniscus effects, which arose during the layer-by-layer fabrication of the hydrogel.[75] Despite this offset, the actuator maintained a substantial angular separation of around 60° between the three states (RCP, CPL-off, and LCP), providing sufficient angular contrast to resolve three macroscopic states. The corresponding average actuation speed was approximately 4° min$^{-1}$. Collectively, these findings established the feasibility of a CPL-responsive soft actuator based on a chiral MXene-embedded hydrogel. This system functioned as a CPL-based ternary actuator, exhibiting three distinct macroscopic actuation modes in response to three different input states: downward bending under RCP, upward bending under LCP, and flattening in the absence of CPL. These results demonstrate ternary, chirality-encoded actuation, in contrast to conventional intensity- or wavelength-controlled photoactuators.

By designing the chiral MXene-embedded hydrogel actuator, we confirmed that CPL served as a ternary input to drive distinct macroscopic actuation. However, while the actuator demonstrated handedness-dependent actuation in response to CPL irradiation, its rod-like geometry inherently restricted its deformations to relatively simple bending behaviors. Based on this established CPL-responsiveness, we next sought to design a more complex architecture whose overall shape deformation could be programmed through spatially controlled CPL irradiation. To this end, we fabricated a square hydrogel construct with a hollow center using PNIPAM-based hydrogel embedded with chiral MXene. Importantly, similar to the actuator described above, the outer frame of the construct was composed of *D*-MXene hydrogel, while the inner frame was composed of *L*-MXene hydrogel **(Figure 5A)**. The outer and inner layers were prepared with equal concentrations of chiral MXene, distinguished solely by their handedness, as validated by optical density measurements. Based on the CPL-sensitive photothermal conversion, we postulated that irradiation of individual sides of the square hydrogel with either LCP or RCP induces inward or outward bending, respectively. Compared to the previous actuator, the key point of this square hydrogel construct lay in its programmability. By applying three different input states (RCP, LCP, or CPL-off) at various locations within a single construct, we anticipated the emergence of a diverse and more complex repertoire of global shape deformations.

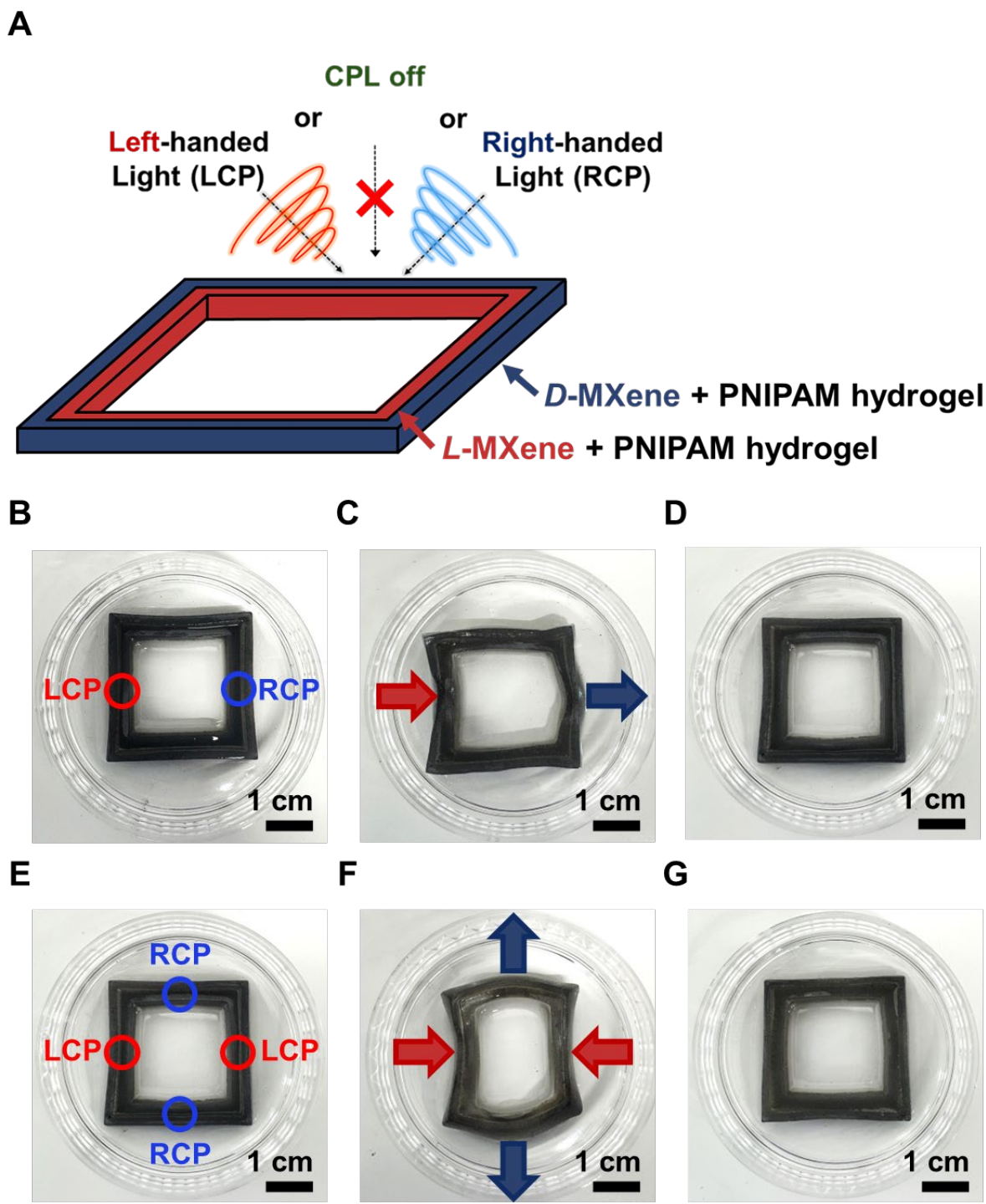

**Figure 5.** CPL-programmable hydrogel architecture using PNIPAM-based hydrogel with chiral MXene. (A) Schematic illustration of the CPL-programmable hydrogel architecture. (B) Photograph of the CPL-programmable hydrogel architecture. Red circles and blue circles indicate regions irradiated with LCP and RCP, respectively. (C) The hydrogel architecture after 20 min of CPL irradiation, showing handedness-dependent deformation. Red arrows and blue arrows denote inward and outward bending, respectively. (D) The hydrogel architecture restored to its original square shape after cooling to ambient temperature. (E) Photograph of the recovered hydrogel architecture and irradiation sites for its CPL-programmable deformation. (F) Programmable deformation of the hydrogel architecture after 20 min of CPL irradiation under the conditions shown in (E). (G) The hydrogel architecture returned to its original square shape after cooling to ambient temperature, confirming reversibility after two consecutive global shape deformations.

To validate the concept of CPL-programmable shape deformation, we irradiated one side of the square hydrogel construct with LCP and the opposite side with RCP for 20 min, while leaving the remaining two sides unexposed **(Figure 5B)**. As a result, the side exposed to LCP exhibited inward bending, driven by the relative contraction of the outer *D*-MXene-containing layer. In contrast, the side exposed to RCP exhibited outward bending, attributed to the greater contraction of the inner *L*-MXene-containing layer. Meanwhile, the two sides without CPL irradiation showed no noticeable deformation **(Figure 5C)**. These observations confirmed that by delivering different input states (RCP, LCP, or CPL-off) to different localized regions of a single construct, programmable shape deformation was achieved. When the deformed construct was cooled to ambient temperature, it returned to its original square configuration, demonstrating that the chiral MXene-based hydrogel construct exhibited reversible shape deformation **(Figure 5D)**.

We next induced more complex structural deformations by irradiating all four sides of the hydrogel construct, which had been restored to its square shape by cooling, with CPL. Specifically, one pair of opposing sides was exposed to LCP, while the remaining pair was exposed to RCP for 20 min **(Figure 5E)**. As a result, the sides irradiated with LCP exhibited inward bending, showing arc-like shapes with central angles of approximately 47.3°. The sides irradiated with RCP exhibited outward bending, also showing arc-like shapes with central angles of approximately 48.4° **(Figure 5F)**. Consequently, the original square construct with a 1:1 aspect ratio and four-fold symmetry was transformed into an asymmetric elongated shape with reduced two-fold symmetry and an aspect ratio of approximately 1.5:1. Notably, this CPL-driven programmable macroscopic symmetry breaking and shape reconfiguration were reversible, as the construct returned to its original square geometry upon cooling to ambient temperature **(Figure 5G)**. These results showed that local CPL handedness can be used as an addressable input for site-specific deformation within a single construct.

**Conclusions**

In this study, we established a chiral $Ti_3C_2T_x$ MXene platform that translates molecular chirality into macroscopic optical control through supramolecular engineering. The resulting material exhibits exceptionally broadband circular dichroism spanning UV to NIR, originating from the synergistic interplay between vertically aligned chiral ligands and MXene's intrinsic optoelectronic properties. This supramolecular chirality enabled CPL-responsive photothermal conversion with a ~30% difference in photothermal heating between matched and mismatched CPL, sufficient contrast to drive distinct conformational changes in thermoresponsive hydrogels.

Most significantly, this work establishes a paradigm for ternary optical logic in soft matter, utilizing light handedness as the primary control variable. The demonstrated CPL-responsive actuator and programmable hydrogel architecture represent a realization of multidirectional, site-specific deformation through chirality-encoding. By combining the broadband chiroptical MXene platform with an optical ternary input knob, we enable spatial multiplexing and addressable actuation that are inaccessible to conventional binary photosystems. Collectively, our findings highlight chiral 2D materials as promising platforms for optically programmable matter governed by polarization logic, going beyond conventional binary photoactuation. The chiral nanopainting strategy demonstrated here is generalizable to other 2D materials, potentially yielding a library of polarization-responsive nanosheets with tunable spectral ranges and response kinetics. This molecular-to-macroscopic design framework—where supramolecular chirality dictates optical selectivity, and optical selectivity commands mechanical deformation—provides a versatile blueprint for future chirality-encoded technologies in soft robotics, bionics, adaptive photonics, and reconfigurable architectures.

**Experimental Section**

**Synthesis of Ti$_3$C$_2$T$_x$ MXene.**

Ti$_3$C$_2$T$_x$ MXenes were synthesized according to reported procedures.[76] A mixture of 60 g of TiC (Yee Young Cerachem, 99.5%), 30 g of Ti (Yee Young Cerachem, 99.5%), and 30 g of Al (Yee Young Cerachem, 99%) powders was ball-milled for 24 h. The milled powder was then sintered at 1435 °C for 2 h. To remove any residues generated during the sintering process, the resulting products were washed twice with 300 mL of 9 M HCl (Sigma-Aldrich, diluted from 37% HCl) and dried at 80 °C for 6 h. A mixture of 1 g of Ti$_3$AlC$_2$, 2 g of LiF (Aldrich, ≥99.99%), and 20 mL of 6 M HCl solution was prepared and stirred magnetically at room temperature for 24 h. After etching, the multilayer MXene solution was centrifuged, the clear supernatant was discarded, and the sediment was redispersed in deionized water; this process was repeated at least four times. The final black supernatant, containing single-layered MXene flakes, was obtained after shaking for 1 h and was then centrifuged at 3500 rpm for 5 min to yield a Ti$_3$C$_2$T$_x$ MXene dispersion in deionized water at a concentration of 10 mg mL$^{-1}$.

**Synthesis of APTES-treated MXene.**

Ti$_3$C$_2$T$_x$ MXene (100 mg) was centrifuged at 17,000 rpm for 20 min in a 40 mL centrifuge tube to precipitate the material. After centrifugation, 5 mL of (3-aminopropyl)triethoxysilane (APTES; Sigma-Aldrich, 99%) solution was added to the precipitate using a syringe, followed by the addition of a trace amount of ethanol. A stirring bar was then placed into the tube, and the mixture was shaken for 180 s to ensure thorough mixing. The tube was stirred at room temperature for 24 h. After completion of the reaction, the stirring bar was removed, and the mixture was subjected to sonication for 3 h with ice added to prevent an increase in temperature. Following sonication, the mixture was centrifuged again at 10,000 rpm for 15 min to precipitate the MXene, and the supernatant containing the APTES solution was discarded. The precipitated

MXene was then dispersed in 10 mL of deionized water to obtain a final concentration of 10 mg mL$^{-1}$.

**Synthesis of Chiral MXene.**

Chiral MXene was prepared according to the previously reported chiral nanopainting method adapted for Ti$_3$C$_2$T$_x$ MXene.[5] 99.1 mg of *L*-phenylalanine (Sigma-Aldrich, ≥99.0%) or *D*-phenylalanine (Sigma-Aldrich, ≥98%) was dissolved in 30 mL of 0.1 M 2-(*N*-morpholino)ethanesulfonic acid (MES; Sigma-Aldrich, ≥99.5%) buffer (pH 6.0). For preparing achiral *DL*-MXene, 99.1 mg of racemic mixture of *L*- and *D*-phenylalanine was used. 230 mg of *N*-(3-(dimethylamino)propyl)-*N*′-ethylcarbodiimide hydrochloride (EDC; Sigma-Aldrich, ≥98.0%) and 138 mg of *N*-hydroxysuccinimide (NHS; Sigma-Aldrich, 98%) were each dissolved separately in 5 mL of 0.1 M MES buffer (pH 6.0). 5 mL of the EDC solution and 5 mL of the NHS solution were added to 30 mL of the *L*- or *D*-phenylalanine solution, sequentially. The mixed solution was stirred for 15 min. Then, 15 μL of the APTES-MXene solution (10 mg mL$^{-1}$) was added to the mixed solution and stirred for 3 h. After 3 h of the reaction, the resulting solution was centrifuged (5,000 × g, 20 min) using an Amicon® Ultra Centrifugal Filter (Millipore, 10 kDa MWCO). The black chiral MXene precipitate remaining within the filter was dispersed in deionized water and centrifuged once again using the Amicon® Ultra Centrifugal Filter. The collected chiral MXene precipitate was re-suspended in deionized water at desired concentrations for further use.

**Synthesis of the PNIPAM-based Hydrogels with Chiral MXene.**

226.3 mg of *N*-isopropylacrylamide (NIPAM; Sigma-Aldrich, 97%), 16 mg of *N*,*N*′-methylenebisacrylamide (NMBA; Sigma-Aldrich, ≥99.5%), and 60 mg of potassium persulfate (KPS; Sigma-Aldrich, ≥99.0%) were each dissolved separately in 1 mL of N$_2$-bubbled deionized water to remove dissolved $O_2$. Chiral MXene was dispersed in $N_2$-bubbled deionized water to a concentration of 0.354 mg mL$^{-1}$, and 0.5 mL of the dispersion was prepared for subsequent use. 650 μL of the NIPAM solution, 251 μL of the NMBA solution, and 2 μL of *N*,*N*,*N*′,*N*′-tetramethylethylenediamine (TEMED; Sigma-Aldrich, ≥99.0%) were added to the chiral MXene dispersion. 264 μL of the KPS solution was then added to the mixture to initiate polymerization of NIPAM, followed by 1 min of sonication for thorough mixing. After 3 h of polymerization, the resulting hydrogel was washed with deionized water twice.

**Synthesis of the CPL-responsive Hydrogel Actuator using Chiral MXene.**

Prior to fabricating the rod-like CPL-responsive hydrogel actuator, a polydimethylsiloxane (PDMS) mold for the hydrogel was prepared. To make the PDMS mold, SYLGARD$^{TM}$ 184 Silicone Elastomer Base and Curing Agent (Dow) were combined at a 10:1 ratio by mass and cured at 60 °C overnight. The resulting PDMS slab contained a central trench (0.5 cm width × 2.5 cm length × 0.8 cm depth), which served as a mold for hydrogel formation.

226.3 mg of NIPAM, 16 mg of NMBA, and 60 mg of KPS were each dissolved separately in 1 mL of $N_2$-bubbled deionized water. *L*- and *D*-MXene were each dispersed separately in $N_2$-bubbled deionized water to a concentration of 1 mg mL$^{-1}$, and 70.6 μL of each dispersion was prepared for subsequent use. The concentrations of the *L*- and *D*-MXene dispersions were verified to be identical by measuring their optical density at 520 nm.

For the bottom *L*-MXene-containing layer, 100 μL of the NIPAM solution, 38.6 μL of the NMBA solution, and 0.3 μL of TEMED were added to the *L*-MXene dispersion. 40.6 μL of the KPS solution was then introduced, and the mixture was subjected to sonication for 2 min to ensure complete mixing. After the sonication, 200 μL of the resulting mixture was injected into the PDMS mold.

For the middle PNIPAM layer, 100 μL of the NIPAM solution, 38.6 μL of the NMBA solution, and 0.3 μL of TEMED were added to 70.6 μL of N$_2$-bubbled deionized water. 40.6 μL of the KPS solution was then introduced, and the mixture was subjected to sonication for 2 min to ensure complete mixing. After 15 min of polymerization of the *L*-MXene-containing mixture in the mold, 100 μL of the prepared PNIPAM solution was injected into the PDMS mold to form another layer above the bottom *L*-MXene-containing layer.

For the top *D*-MXene-containing layer, 100 μL of the NIPAM solution, 38.6 μL of the NMBA solution, and 0.3 μL of TEMED were added to the *D*-MXene dispersion. 40.6 μL of the KPS solution was then introduced, and the mixture was subjected to sonication for 2 min to ensure complete mixing. After 15 min of polymerization of the middle achiral PNIPAM mixture in the mold, 200 μL of the resulting mixture with *D*-MXene was injected into the PDMS mold. The layered hydrogel was left at room temperature for 3 h to allow complete polymerization. After 3 h of polymerization, the resulting hydrogel was washed with deionized water twice.

**Synthesis of the CPL-programmable Hydrogel Architecture using Chiral MXene.**

Prior to fabricating the square CPL-programmable hydrogel architecture, a PDMS mold for the hydrogel was prepared. SYLGARD™ 184 Silicone Elastomer Base and Curing Agent were combined at a 10:1 ratio by mass and cured at 60 °C overnight to make a PDMS slab. A square cavity (3 cm × 3 cm, depth 1 cm) was fabricated at the center of the slab and used as a mold for hydrogel formation. Separately, cubic structures with edge lengths of 2.5 cm and 2.0 cm, respectively, were fabricated from polyvinylidene fluoride (PVDF) for subsequent hydrogel preparation.

226.3 mg of NIPAM, 16 mg of NMBA, and 60 mg of KPS were each dissolved separately in 1 mL of N$_2$-bubbled deionized water. *L*- and *D*-MXene were each dispersed separately in $N_2$-bubbled deionized water to a concentration of 1 mg mL$^{-1}$, and 459 μL of each dispersion was prepared for subsequent use. The concentrations of the *L*- and *D*-MXene dispersions were verified to be identical by measuring their optical density at 520 nm.

For the outer *D*-MXene-containing layer, 650 μL of the NIPAM solution, 251 μL of the NMBA solution, and 2 μL of TEMED were added to the *D*-MXene dispersion. Subsequently, 459 μL of the KPS solution was introduced, and the mixture was subjected to sonication for 2 min to ensure complete mixing. After the sonication, the PVDF cube with an edge length of 2.5 cm was positioned at the center of the square PDMS cavity (3 cm per side). Subsequently, 1.375 mL of the prepared mixture with *D*-MXene was injected into the square channel formed between the PDMS mold and the PVDF cube. Given that the basal area of the square channel was 2.75 cm$^2$, the injected mixture filled the channel to a height of 0.5 cm.

For the inner *L*-MXene-containing layer, 650 μL of the NIPAM solution, 251 μL of the NMBA solution, and 2 μL of TEMED were added to the *L*-MXene dispersion. Subsequently, 459 μL of the KPS solution was introduced, and the mixture was subjected to sonication for 2 min to ensure complete mixing. After the sonication, the PVDF cube at the center of the cavity was carefully removed and the PVDF cube with an edge length of 2 cm was positioned at the center of the square cavity (2.5 cm per side) surrounded by the outer *D*-MXene-containing hydrogel. 1.125 mL of the prepared mixture with *L*-MXene was injected into the square channel formed between the outer hydrogel and the PVDF cube. Given that the basal area of the square channel was 2.25 cm$^2$, the injected mixture filled the channel to a height of 0.5 cm. The layered hydrogel was left at room temperature for 3 h to allow complete polymerization. After 3 h of polymerization, the PVDF cube at the center of the cavity was carefully removed and the resulting hydrogel was washed with deionized water twice.

**Characterization of Chiral MXene.**

UV-vis-NIR absorption and CD spectra measurements were performed using a J−1700 instrument (Jasco). SEM imaging was carried out using a Magellan 400 (FEI), and the samples were prepared by drop-casting the dispersions onto Anodisc membrane filters (Cytiva). TEM imaging was carried out using a Talos F200X (Thermo Scientific) with copper-coated carbon TEM grids. XPS analysis was performed with a K-Alpha+ (Thermo Scientific).

**CPL-dependent photothermal conversion experiments.**

To investigate the handedness-dependent photothermal conversion of chiral MXene, the optical system was established using an NIR laser source (780 nm, 1 W) with a beam diameter of 0.5 cm, a linear polarizer (Thorlabs) and a quarter-wave plate (Thorlabs) appropriate for a 780 nm laser. Initial unpolarized light from the NIR laser was transformed into linearly polarized light by passing through the linear polarizer, and the linearly polarized light was transformed into LCP or RCP by passing through the rotated quarter-wave plate (45° or -45°).

CPL irradiation was applied to 0.2 mL of the chiral MXene aqueous dispersion. The dispersion concentration and irradiation duration are provided in the supporting information. The temperature of the dispersion was measured using a TE-SQ1 (Thermal Expert).

**CPL-responsive actuation and deformation experiments.**

To investigate the actuation of the rod-like CPL-responsive hydrogel actuator and the deformation of the square CPL-programmable hydrogel architecture, the optical system was established using a VIS laser source (520 nm, 1 W) with a beam diameter of 0.5 cm, a neutral density filter with an optical density of 0.5 (Edmund Optics), a linear polarizer (Thorlabs) and a quarter-wave plate (Thorlabs) appropriate for a 520 nm laser. Initial unpolarized light from the VIS laser passed through the neutral density filter to adjust the input intensity. The light was then transformed into linearly polarized light by passing through the linear polarizer, and the linearly polarized light was transformed into LCP or RCP by passing through the rotated quarter-wave plate (45° or -45°).

CPL irradiation was applied to the center of the hydrogel actuator for the CPL-responsive hydrogel actuator experiments, and to the center of each side for the CPL-programmable hydrogel architecture experiments. The irradiation duration for each case is provided in the main text.

## Supporting Information

**Supporting Information Available:** Additional figures including TEM images, XPS spectra, CD spectra, and images of the CPL-responsive actuator. (DOCX)

## Author Contributions

J. Y. conceived the study. W. J., Y. L., and K. H. P. synthesized and analyzed the chiral MXenes using TEM, SEM, XPS, and CD. D.L. conducted the molecular docking simulations. W. J. and K. H. P. synthesized and analyzed the PNIPAM-based chiral hydrogel with the chiral MXene. W. J. fabricated the CPL-responsive soft actuator and the CPL-programmable hydrogel architecture. W.J., D.L., and J.Y. analyzed the results and wrote the manuscript with comments from all authors.

## Conflict of Interest

The authors declare no conflict of interest.

## Acknowledgements

**Data Availability Statement**

The data that support the findings of this study are available in the supporting information of this article.